\documentclass[iop]{emulateapj}
\usepackage[squaren, Gray, cdot]{SIunits}
\usepackage{graphicx}
\usepackage[caption=false]{subfig}
\usepackage{amsmath}
\usepackage{color}
\def\mearth{M_\oplus}
\def\rearth{R_\oplus}

\slugcomment{Accepted for publication in The Astrophysical Journal Letters}

\shorttitle{Possible Internal Structures and Compositions of Proxima Centauri b}
\shortauthors{Brugger et al.}

\begin{document}

\title{Possible Internal Structures and Compositions of Proxima Centauri b}

\author{
B. Brugger\altaffilmark{1},
O. Mousis\altaffilmark{1},
M. Deleuil\altaffilmark{1},
and J. I. Lunine\altaffilmark{2}
}

\altaffiltext{1}{Aix Marseille Univ, CNRS, LAM, Laboratoire d'Astrophysique de Marseille, Marseille, France (\email{bastien.brugger@lam.fr})}
\altaffiltext{2}{Department of Astronomy and Carl Sagan Institute, Space Sciences Building Cornell University,  Ithaca, NY 14853, USA}

\begin{abstract}

We explore the possible Proxima Centauri b's interiors assuming the planet belongs to the class of dense solid planets (rocky with possible addition of water) and derive the corresponding radii. To do so, we use an internal structure model that computes the radius of the planet along with the locations of the different layers of materials, assuming that its mass and bulk composition are known. Lacking detailed elementary abundances of the host star to constrain the planet's composition, we base our model on solar system values. We restrained the simulations to the case of solid planets without massive atmospheres. With these assumptions, the possible radius of Proxima Centauri b spans the 0.94--1.40 $\rearth$ range. The minimum value is obtained considering a 1.10 $\mearth$ Mercury-like planet with a 65\% core mass fraction, whereas the highest radius is reached for 1.46 $\mearth$ with 50\% water in mass, constituting an ocean planet. Although this range of radii still allows very different planet compositions, it helps characterizing many aspects of Proxima Centauri b, such as the formation conditions of the system or the current amount of water on the planet. This work can also help ruling out future measurements of the planet's radius that would be physically incompatible with a solid planetary body.

\end{abstract}


\section{Introduction}
\label{sec:1}

The recent discovery of a planet orbiting the Sun's nearest star, the red dwarf Proxima Centauri (Anglada-Escud\'e et al. 2016), has brought the search for extraterrestrial life on exoplanets to the gates of our solar system. The mass of the planet Proxima Centauri b (hereafter Proxima b), ranging close to that of the Earth, and the fact that it follows an orbit within a temperate zone (Anglada-Escud\'e et al. 2016), make real the possibility that has been found an Earth analog (or at least a habitable planet) within our nearest neighborhood. This planet was discovered from reanalysis of previous radial velocity measurements, which yield the minimum mass $m$~sin~$i$, where $i$ is the (unknown) orbital inclination angle. The reported minimum mass is 1.10--1.46 Earth masses ($\mearth$). We adopt this mass range here in our modeling (setting sin~$i = 1$) because it is astrobiologically interesting, and allows us to use equations of state well-tested for the Earth. Proxima b orbits its host star with a period of 11.2~days, corresponding to a semi-major axis distance of 0.05~AU. Unlike many other exoplanets discovered so far, this short distance does not imply a high surface temperature for the planet. Proxima Centauri being a red dwarf, its luminosity is only 0.15\% of that emitted by the Sun, and its effective temperature is 3,050~K. Therefore, a planet located at 0.05 AU from Proxima Centauri has an equilibrium temperature of only $\sim$234~K (Anglada-Escud\'e et al. 2016). This temperature is close to the melting point of water if one assumes that Proxima b is surrounded by an atmosphere with a surface pressure of one bar, implying that the planet lies within the habitable zone of its host star.

Because no transit signal has been observed for Proxima b, its radius is unknown. Chances for having this quantity determined in the future remain low since Anglada-Escud\'e et al. (2016) predict a geometric probability of transit of only 1.5\% from their data. Here we explore the possible Proxima b's interiors assuming the planet belongs to the class of dense solid planets  (rocky with possible addition of water) and derive the corresponding radii. To do so, we use an internal structure model that computes the radius of the planet along with the locations of the different layers of materials, assuming that its mass and bulk composition are known. We excluded from this study cases where Proxima b could harbor a thick atmosphere, unlike the Earth, to focus on planets that are likely to be habitable.

\section{Internal structure}
\label{sec:2}

\subsection{Physical model}
\label{ssec:2.1}

Our model follows the approach described by Sotin et al. (2007). It consists of a one-dimensional model that computes the internal structure and radius of a planet from a given mass and composition. The Earth is taken as a reference for various parameters and constants since its composition and internal structure are well known. Our interior model is composed of the following five fully differentiated layers:

\begin{itemize}
\item[-] a metallic core, composed of a mixture of pure iron and iron alloy (FeS). Here, the double-layered core of the Earth is reduced to a single layer;
\item[-] the lower mantle, made of silicate rocks perovskite and magnesiow\"ustite;
\item[-] the upper mantle made from the same elements but in the form of olivine and enstatite;
\item[-] a high-pressure water ice layer (ices VII and X; Hemley et al. 1987);
\item[-] a layer made of liquid water (equivalent to Earth's oceans).
\end{itemize}

\noindent The sizes and masses of these different layers are variable, allowing our model to simulate planetary compositions varying from terrestrial (fully rocky) planets to ocean planets. Several input parameters must be defined to describe the precise composition of a planet, namely the fraction of iron alloy in the core, the fraction of iron in the mantles (given by Mg\#, a parameter that expresses the differentiation of the planet, defined in Table~\ref{tab:table1}) and the relative distribution of the different types of silicate rocks in the mantles (fixed by the Mg/Si ratio of the body). Here we use the Earth's values by default, listed in Table~\ref{tab:table1}, because the parameters are unknown in the case of Proxima b. Thus, 90\% of the mantle silicates are based on Mg, the remaining 10\% are based on Fe. However, these parameters have a limited influence on a given planet's radius (Sotin et al. 2007, Valencia et al. 2006, 2007a) compared to the core mass fraction (CMF) and water mass fraction (WMF), which determine the positions of the core/lower mantle and upper mantle/water layer boundaries, respectively. Remaining boundaries (lower/upper mantle, water ice/liquid water) are directly computed from the phase diagrams of the respective materials.

\begin{deluxetable}{lcl}
\tablecaption{Compositional parameters and surface conditions of the Earth, used for Proxima Centauri b}
\tablehead{
\colhead{Parameter} & \colhead{Value} & \colhead{Description}
}
\startdata
$f_{alloy}$			& 0.13		& Fraction of FeS alloy in the core			\\
Mg\#				& 0.9		& Mole fraction Mg/(Mg+Fe) in the mantles	\\
Mg/Si				& 0.131		& Mg/Si mole ratio of the body				\\
T$_{surf}$ (K)		& 288		& Surface temperature						\\
P$_{surf}$ (bar)	& 1			& Surface pressure							\\
\enddata
\tablecomments{Compositional values are taken from Sotin et al. (2007).}
\label{tab:table1}
\end{deluxetable}

The internal structure of a planet is governed by the gravitational acceleration $g$, pressure $P$, temperature $T$ and density $\rho$ inside the body. These quantities are computed by solving the corresponding differential equations, completed by the one verifying the planet mass:

\begin{equation}
\dfrac{d g}{d r} = 4 \pi G \rho - \dfrac{2 G m}{r^3},
\label{eq:equation1}
\end{equation}
\begin{equation}
\dfrac{d P}{d r} = - \rho g,
\label{eq:equation2}
\end{equation}
\begin{equation}
\dfrac{d T}{d r} = - g \dfrac{\gamma T}{\Phi},
\label{eq:equation3}
\end{equation}
\begin{equation}
\dfrac{d m}{d r} = 4 \pi r^2 \rho,
\label{eq:equation4}
\end{equation}

\noindent where $r$ is the radius inside the planet, $m$ the mass at a given radius, $G$ is the gravitational constant, and $\gamma$ and $\Phi$ are the Gr\"uneisen and seismic parameters, respectively (see Sotin et al. 2007 and Valencia et al. 2006 for details). The planet's interior is modeled by a one-dimensional spatial grid with fixed precision, ranging from the planet center until beyond its surface, and each physical quantity is computed for every point of the grid. Gravitational acceleration and density are integrated numerically from the planet center, whereas pressure and temperature are integrated from the surface. The model iterates on the solution of Eqs \ref{eq:equation1}--\ref{eq:equation4}, calculating new layer boundaries at each iteration, until it reaches convergence. Convergence is achieved when the planet matches the asked mass, and when the boundary conditions are verified: no central gravitational acceleration, surface pressure and temperature fixed to the given values. In our case, since Proxima b lies within the habitable zone of Proxima Centauri (Anglada-Escud\'e et al. 2016), we use the surface conditions of the Earth (see Table~\ref{tab:table1}). In the following, we assume the presence of a thin Earth-like atmosphere that allows the existence of liquid water on Proxima b. This latter does not have any influence on our computations of the planet radius since its mass only accounts for 0.0001\% of the Earth's mass.

\subsection{Equations of state}
\label{ssec:2.2}

Equations of state (EOS) are needed to compute the density profile of a given planet. Sotin et al. (2007) employed two well known EOS: the third-order Birch-Murnaghan (BM3) EOS to depict the density of the planet's outer layers (upper mantle and liquid water layer), and the Mie-Gr\"uneisen-Debye (MGD) EOS to compute the density within the core, the lower mantle and the water ice layer. The BM3 EOS is particularly fast for computations, but presents several limitations as the terms neglected in the third-order development may exceed the lower ones for too high pressure values ($>$1.5--3~Mbar; Seager et al. 2007, Valencia et al. 2009) or temperatures higher than the ones encountered in the Earth (Sotin et al. 2007). In contrast, the MGD EOS does not present this limitation and can be used in the core and the lower mantle. In the present case, we opted to replace the MGD EOS by another EOS already used by Valencia et al. (2007a), namely the Vinet EOS. Although its validity range in pressure is similar to that of the BM3 EOS, the Vinet EOS has been shown to be a better fit to experimental data than BM3, and to better extrapolate at high and very high pressures (Hama and Suito 1996, Cohen et al. 2000). Therefore it is well adapted to the modeling of Super-Earths, where such pressures can be encountered. We use the Vinet EOS in the dense layers (the core and the mantles) and keep the BM3 formula when computing the density in the water layers.

\subsection{Ternary diagram}
\label{ssec:2.3}

\begin{figure}
\begin{center}
\includegraphics[width=0.9\columnwidth]{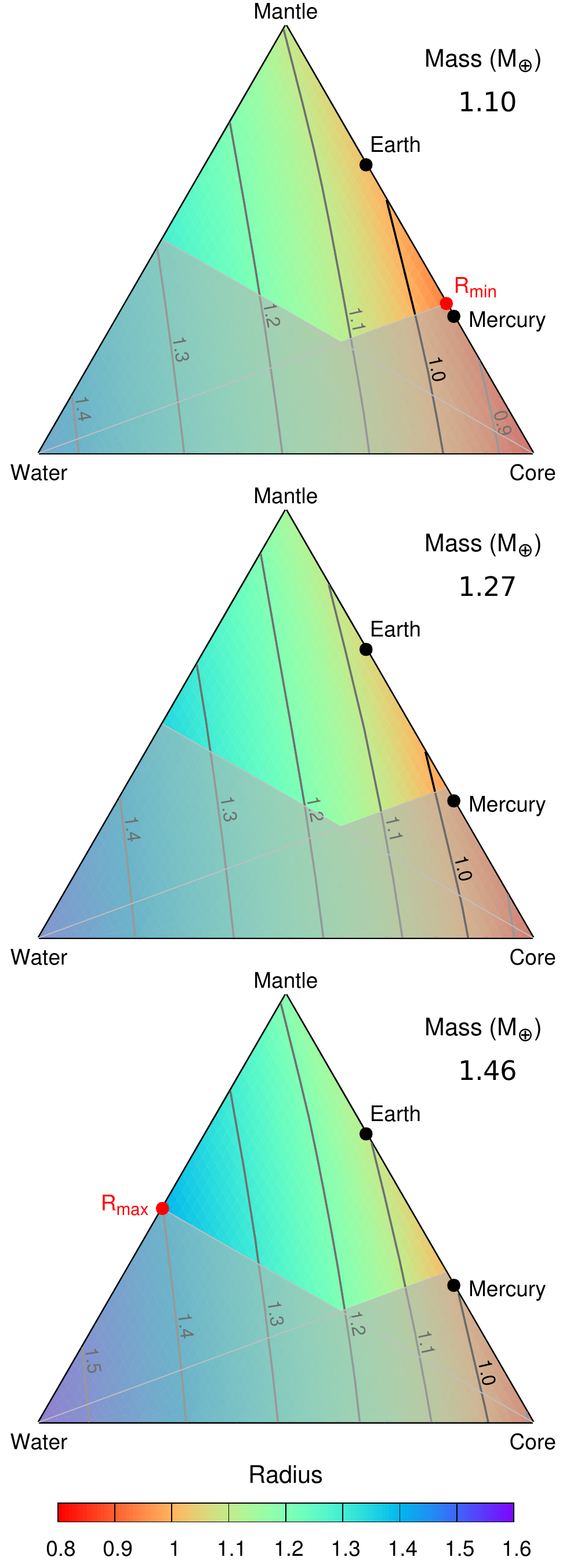}
\caption{Ternary diagrams displaying the investigated compositional parameter space for different $M_P$ values. From top to bottom, diagrams correspond to the minimum, mean and maximum masses inferred for Proxima b. Isoradius curves ranging between 0.8 and 1.6 $\rearth$ are also represented. Compositions (but not their mass nor their radius) of the Earth and Mercury are placed on each diagram (see Section~\ref{ssec:2.3} for the corresponding values of CMF and WMF).}
\label{fig:figure1}
\end{center}
\end{figure}

In the following, because the internal structure of a planet is mainly governed by the CMF and WMF parameters, its ``composition'' refers to the pair (CMF,WMF). For instance, an Earth-like composition corresponds to (CMF,WMF)=(0.325,0) (Stacey 1992), and a Mercury-like composition to (CMF,WMF)=(0.68,0) (Schubert et al. 1988, Harder and Schubert 2001). The parameter space allowed by the variations of the CMF and the WMF are represented by ternary diagrams, displayed by Figure \ref{fig:figure1}. Such diagrams present three preferential directions, each of them corresponding to a key compound of our model (core, mantle and water). Each point on this diagram corresponds to a unique composition: the corners match planets fully composed of the corresponding compound (core, mantle or water). In contrast, a planet located on one side does not contain any fraction of the material indicated on the opposite corner (as illustrated by the cases of the Earth and Mercury, in which the water fraction is considered zero). Each ternary diagram, corresponding to a specified planet mass, shows the range of plausible radii achieved by the planet by spanning all the possible compositions. In other words, each point on the ternary diagram corresponds to a specific set of composition and radius (the iteration method explained in Section~\ref{ssec:2.1} is processed once for each point of the diagram).

However, some areas in the diagram correspond to planet compositions which cannot physically exist as planetary bodies. To identify these areas, we use the work of Valencia et al. (2007b) who show that two limits can be drawn from hypotheses on the solar system formation: (i) there is a maximum allowed value for the CMF ($\sim 65\%$) since the proportions of metal and silicates are fixed by the Fe/Si ratio of the solar nebula, and (ii) a planet cannot present a water proportion higher than 77\% according to measurements on cometary compositions. In this work we place an even stronger limit on the WMF by lowering the maximum value to 50\%, the value typically derived from interior models of large icy satellites such as Titan (Tobie et al. 2006). The two areas excluded by these limits are shaded in the ternary diagram. 

Our approach allows us to compute a range of possible values for the radius of Proxima b, under the condition that the planet is solid. Any future measurement of the radius falling out of this range will imply the following for Proxima b:
\begin{itemize}
\item if lower, Proxima b would present a CMF higher than 65\%, meaning that the planet would have undergone an important event not considered in formation scenarios, as it is the case in the solar system. The causes proposed for Mercury's 68\% CMF are a strong mantle evaporation (Cameron 1985) or a giant impact (Benz et al. 1988, 2007, Harder and Schubert 2001);
\item if the measured radius is higher, either (i) Proxima b holds more water than any body in the solar system, or (ii) it is made of materials lighter than those considered here (most probably a significant H/He atmosphere).
\end{itemize}
\noindent The range of radii provided by our model covers the case of a solid Proxima b solely composed of metal, silicates and/or water, in agreement with the solar system formation conditions.

\section{Results}
\label{sec:3}

Figure~\ref{fig:figure1} shows the results of our computations in the compositional parameter space allowed for Proxima b. We have explored three different planet masses $M_P$, namely 1.10, 1.27 and 1.46 $\mearth$, corresponding to the minimum, mean and maximum measured values, respectively. The ternary diagrams display the range of plausible radii for each adopted $M_P$ value. The minimum radius (0.94 $\rearth$) is determined from the adoption of the minimum $M_P$ value (top of Figure~\ref{fig:figure1}), with 65\% of the planet's mass located in the core and the remaining 35\% as part of the mantle (i.e. no water layer). Conversely, the maximum radius (1.40 $\rearth$) is derived from the adoption of the maximum $M_P$ value (bottom of Figure~\ref{fig:figure1}). In this case, the corresponding composition is 50\% of the planet's mass in the form of water and the remaining 50\% in the mantle (i.e. no core). These computations take into account the limitation proposed by Valencia et al. (2007b) for the CMF, and our limitation for the WMF. If these limitations are neglected, the derived range of radii for Proxima b would be 0.82--1.55 $\rearth$, with 100\% core and 100\% water compositions for 1.10 and 1.46 $\mearth$, respectively.

Figure~\ref{fig:figure1} also displays lines, named isoradius curves, on which the computed planet radius is constant. It is interesting to compare the behavior of the 1 $\rearth$ isoradius curve between the three ternary diagrams. For $M_P = 1.10$ $\mearth$, a 1 $\rearth$ radius is easily achievable for planet compositions with larger CMF than that of the Earth. Interestingly, half of the curve is located in the exclusion zone. When increasing the mass to 1.27 $\mearth$, the curve shifts to the right of the diagram, as a result of the increase of the planet's density and CMF value. In this case, only a small fraction of the 1 $\rearth$ curve lays outside the exclusion zone. This trend becomes obvious with $M_P = 1.46$ $\mearth$, since the 1 $\rearth$ curve is completely confined in the exclusion zone. This result implies that if $M_P$ $\ge$ 1.46 $\mearth$, Proxima b cannot present a radius of 1 $\rearth$ (or less), unless it (i) has undergone a particular event that gave it a CMF higher than 65\% (as for Mercury, see Section~\ref{ssec:2.3}), or (ii) is made from elements heavier than iron, which is unlikely.

The range of radii obtained from our computations can be directly compared to the radii measurements of other exoplanets orbiting around different red dwarfs. Using the online tools \textit{exoplanets.eu} (Schneider et al. 2011) and \textit{exoplanets.org} (Han et al. 2014), we listed 30 planets orbiting red dwarfs (with an effective temperature up to 3,750~K) ranging from 0.57 to 12 $\rearth$. This list shows that small planets orbiting red dwarfs seem to be quite usual, since $\sim$57\% of the listed bodies have radii $<$ 1.5 $\rearth$ (close to the upper value inferred for Proxima b), and $\sim$77\% present radii $<$ 2 $\rearth$. Among these planets, GJ 1132 b is the only one presenting determinations of both mass (1.62$\pm$0.55 $\mearth$) and radius (1.16$\pm$0.11 $\rearth$) (Berta-Thompson et al. 2015). For these measurements, we computed the range of possible compositions allowed by the errors on GJ 1132 b's mass and radius. Our model yields a maximum CMF of 96\% for the densest composition (2.17 $\mearth$ and 1.05 $\rearth$, giving a mean density of 10.4 g/cm$^3$), opposed to a maximum WMF of 67\% for the lightest composition (1.07 $\mearth$ with 1.27 $\rearth$, i.e. a mean density of 2.88 g/cm$^3$). In comparison, the range of mean density inferred for Proxima b is 2.21--9.69 g/cm$^3$, which is comparable but slightly lower. Earth's mean density is 5.51 g/cm$^3$, thus both planets could be terrestrial. But unlike Proxima b, GJ 1132 b is probably not habitable. Because of a low orbital distance (0.014 AU), its equilibrium temperature lays within the ~400--600 K range (Berta-Thompson et al. 2015).

\begin{figure}
\begin{center}
\includegraphics[width=1.05\columnwidth]{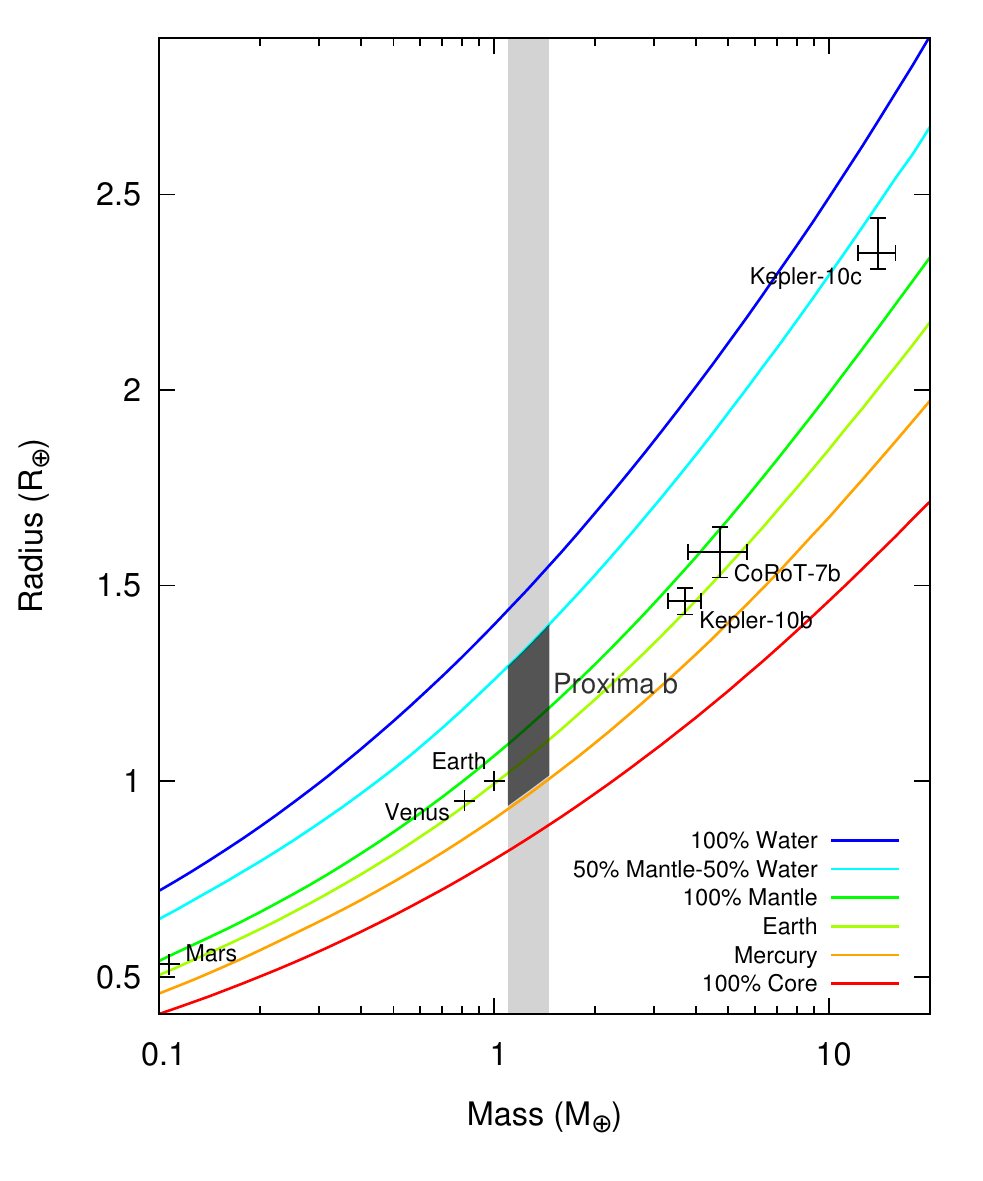}
\caption{Mass-radius diagram showing the curves for different compositions from 100\% core to 100\% water. The light grey and dark grey vertical areas correspond to the range of measured masses for Proxima b (1.10--1.46 $\mearth$) and the range of possible planet radii computed by our model, respectively.}
\label{fig:figure2}
\end{center}
\end{figure}

Figure~\ref{fig:figure2} represents the positions of Proxima b and some other known exoplanets on a mass-radius diagram. The Figure shows that Proxima b is closer to the Earth and Venus than some well known Super-Earths like CoRoT-7b or Kepler-10b, even when considering the uncertainties on its mass and the large range of radii allowed by the model. The mass-radius diagram is a complementary way of representing the possible compositions of an exoplanet. However, each point of the mass-radius diagram yields an infinite number of compositions, that are represented as the isoradius curves on the ternary diagram.

\section{Discussion and conclusions}
\label{sec:4}

Assuming that Proxima b is solid, we find that its radius is in the $\sim$0.94--1.40 $\rearth$ range. The lower value corresponds to a planet made from 65\% iron located in the core and 35\% silicate present in the mantle, meaning that the planet surface is rocky, possibly surrounded by a thin atmosphere. The upper value corresponds to a planet made from 50\% silicate present in the mantle and 50\% water located in an outer layer, mainly constituted of high-pressure ices surrounded with a $\sim$200 km deep liquid ocean (6\% of the total water mass). Proxima b may also present a radius outside the aforementioned range, implying that it would be made from materials others than iron, silicates or water (e.g. if it harbors a thick atmosphere of H/He), or that particular formation and/or evolution conditions would have increased its CMF to a value higher than 65\%.

Our model also helps placing constraints on the planet's formation conditions. Four different formation scenarios have been proposed for Proxima b, with only one case leading to a planet completely dry (planet formation on its current orbit via pebble accretion). The dry planet case corresponds here to WMF = 0 and this limitation reduces the allowed space of compositions to only one side of the ternary diagram shown by Figure~\ref{fig:figure1}. The range of radii for a completely dry Proxima b then becomes 0.94--1.19~$\rearth$. Note that any future measurement of the radius in this range does not necessarily mean that the planet is dry. The other formation scenarios (in situ accretion of planetesimals diffused from the outer parts of the system, or formation of a planetary body/planetary embryos beyond the snowline with following migration to the current orbit) predict a planet containing a significant amount of water and/or volatiles. If we assume this amount to be at least 10\% of the planet's mass (about the WMF of Europa; Sohl et al. 2002), then the smallest achievable planet radius becomes 1.02 $\rearth$ (with $M_P$ = 1.10 $\mearth$, and (CMF,WMF)=(0.59,0.1)). These cases almost cover the full ternary diagram, the band with WMF $<$ 10\% being excluded.

Investigations of the irradiation conditions of Proxima b show that the loss of the planet's water inventory is an open question (Ribas et al. 2016, submitted). This implies that the large possible range of WMF we assumed is still relevant.

The limitations we considered to disregard some regions of the ternary diagram are entirely based on solar system values. These assumptions could be irrelevant for the Proxima Centauri system, since the star is a red dwarf with metallicity [Fe/H] = 0.21. Complementary data on the Proxima Centauri system is of prime importance to better characterize Proxima b. In particular, high resolution spectroscopy of the host star and the determination of its Mg/Si and Fe/Si ratios would provide further constraints on the planet's interior. More generally, a higher precision on future measurements of Proxima b is essential to reduce the compositional variations allowed for this planet.

\acknowledgements
B.B. and O.M. acknowledge support from the A*MIDEX project (n\textsuperscript{o} ANR-11-IDEX-0001-02) funded by the ``Investissements d'Avenir'' French Government program, managed by the French National Research Agency (ANR). O.M. also acknowledges support from CNES. J.I.L. is supported by the James Webb Space Telescope Project.

\end{document}